\documentclass[twocolumn]{aastex63}

\newcommand{\degs}{\mbox{$^{\circ}$}}

\newcommand{\axaf}{\mbox{\em Chandra\/}}

\usepackage{graphicx}
\usepackage{hyperref}
\usepackage{color}

\usepackage{lineno}
\modulolinenumbers[1]

\received{\today 2021}
\revised{\today}
\accepted{TBD}
\submitjournal{ApJ}

\shorttitle{X-ray Structure of the Lensed AGN MGB2016+112}
\shortauthors{Schwartz, Spingola \& Barnacka}
\graphicspath{{./}{figures/}}

\begin{document}

\title{Resolving Complex Inner X-ray Structure of the Gravitationaly Lensed AGN MGB2016+112}


\correspondingauthor{Daniel Schwartz}
\email{das@cfa.harvard.edu}

\author[0000-0001-8252-4753]{Daniel Schwartz}
\affiliation{Smithsonian Astrophysical Observatory, Cambridge, MA 02138, USA}

\author[0000-0002-2231-6861]{Cristiana Spingola}
\affiliation{INAF $-$ Istituto di Radioastronomia, Via Gobetti 101, I$-$40129, Bologna, Italy}

 \author[0000-0001-5655-4158]{Anna Barnacka}
\affiliation{Smithsonian Astrophysical Observatory, Cambridge, MA 02138, USA}
\affiliation{Astronomical Observatory, Jagiellonian University, ul. Orla 171, 30-244 Cracow, Poland}

\begin{abstract}
We use a Chandra X-ray observation of the gravitationally lensed system MGB2016+112 at z=3.273  to elucidate presence of at least two X-ray sources. We find that these sources are consistent with the VLBI components measured by \citet{Spingola19}, which are separated by $\sim 200$ pc. Their intrinsic 0.5 -- 7 keV  source frame luminosities are 2.6$\times$10$^{43}$ and 4.2$\times$10$^{44}$ erg s$^{-1}$. Most likely  this system contains a dual active galactic nucleus (AGN), but we possibly are detecting an AGN plus a pc-scale X-ray jet, the latter lying in a region at very high magnification. The quadruply lensed X-ray source is within $\pm$40 pc (1$\sigma$) of its VLBI counterpart. Using a gravitational lens as a telescope, and a novel statistical application, we have achieved  unprecedented accuracy for measuring metric distances at such large redshifts in X-ray astronomy, which is tens of mas if the source is located close to the caustics, while it is of hundreds of mas if the source is in a region at lower magnification. The present demonstration of this approach has implications for future X-ray investigations of large numbers of lensed systems.
\end{abstract}

\keywords{ Strong gravitational lensing --- Black hole physics: supermassive black holes --- Active galactic nuclei: quasars ([HB89] 2016+112) --- X-ray active galactic nuclei --- astrometry}


\section{Introduction} \label{sec:intro}

The X-ray emission from inner regions of active galaxies is a key to our understanding of supermassive black holes growth, mergers, and accretion processes. However, the study of formation of galaxies in the early Universe is limited by the inability of telescopes to detect and resolve inner regions of these faint objects. At high energies, physical constraints limit further  improvements of the resolution of telescopes. As a result, present and future missions will not provide sufficient resolution to resolve inner regions of merging galaxies potentially hosting multiple AGN. Here we use a novel approach of combining gravitational lensing with the capabilities of the Chandra Observatory to study the system MG B2016+112\footnote{While we retain the B1950 notation for the name, we report all coordinates in the J2000 system.}, located at redshift of 3.273 \citep{Lawrence1993}, when the Universe was just less than 2 billion years old. This complex source is magnified by  a factor of $\mu \sim1.5$ (doubly imaged region) and a factor of $\mu \sim300$ (quadruply imaged region) by gravitational lensing due to its proximity  to the caustic of the lens. Such huge magnification is accompanied by spatial amplification that allows us not only to observe a source that would be otherwise too faint, but also to elucidate its structure as well as resolve spectra of individual lensed images of multiple sources. Radio observations reported by \citet{Spingola19} suggested possible dual-AGN-like structures with misaligned jets, and measured relative motions of components. If confirmed, MG~B2016+112 would be the first gravitationally lensed dual AGN discovered to date.

\subsection{Galaxy Formation and Evolution in the Early Universe}

Hierarchical structure formation simulations 
show that galaxies merge, and thus their central black holes should form dual AGNs (separation less than 10 kpc) and physical binary AGNs (separation less than 100 pc; \citealt{Burke-Spolaor14}). The occurrence of binary and dual AGNs has important implications for assessing the time needed for the  final stage of merging of the two supermassive black holes (SMBH). The eventual coalescence of such SMBH binaries is expected to be a major source of gravitational waves detected by the Laser Interferometer Space Antenna (LISA), or by the Pulsar Timing Array (PTA) if M$>$10$^7$M$_{\odot}$ \citep{Sesana2009, DalCanton19, DeGraf20}.

So far the search for dual AGN has mostly been limited to low redshift (z$\ll$1) and large separation ($\gg$~kpc) using a variety of methods. For instance,  the ultra-luminous infrared galaxy NGC 6240 at z=0.024 was the first X-ray source shown to be a dual AGN \citep{Komossa03}.
\citet{Rubinur19} have found low-z candidates among double peaked emission line radio galaxies, but they note that core-jet, rotating disks, and jet-ISM interactions can also give such signatures. 
\citet{Hwang19} have proposed to use variable sources causing astrometric noise in Gaia to uncover sub-kpc AGN using ACS imaging on Hubble. 
\citet{Gross19} found low luminosity dual AGN at z$<0.22$ with both AGN emitting X-rays at separations from 4 to 7 kpc.
\citet{Deane2014} found a triple SMBH system at z=0.39, with two SMBH separated by only 140 pc, using high resolution interferometric radio observations. 
\citet{Connor2019} found tentative evidence for AGN at 11 kpc separation in a merging system at z=6.23, using a 150 ks \axaf\ observation. Those references reflect the state of the art techniques for identifying dual AGN. Contrast to the results presented here, using a 7.77~ks \axaf\ observation. 

\subsection{X-ray Emission from the Inner Regions of Active Galaxies}

It is widely \emph{assumed} that the multi-wavelength non-thermal emission in quasars has a coincident origin near the central SMBH.  
However, X-ray observations at redshifts greater than a few tenths allow access only to kpc scales, thus providing 1000 times worse resolution as compared to radio observations. These technical limitations prohibit the direct study of the origin of X-ray emission from AGN and their relativistic jets. 
There is much more to learn about the origin and structure of jets if we could probe in greater spatial detail in all wave-bands. We know that the origin of variable emission need not be within the pc-scale core.  

In the nearby source M87, \axaf\ imaging resolved a huge flare in the HST-1 knot 60~pc from the core \citep{Harris03}. This knot dominated the X-ray emission from the nucleus for more than 4 years.  
\citet{Marshall10} and \citet{Hardcastle16} suggested X-ray flaring in a knot of the Pic A jet, tens of kpc from the nucleus.
\citet{Jones20} report that 20\% of the X-rays, including an Fe K line, are more than 1 kpc from the nucleus in the nearby Compton thick AGN NGC 7212. Similar kpc-scale X-rays are also seen in several other nearby AGN \citep[e.g.,][]{Arevalo14, Bauer15, Fabbiano17}. 

\subsection{Gravitational Lenses as High Resolution Telescopes}

Gravitational lensing is the most powerful tool to amplify spatially the images of cosmologically distant sources \citep{Barnacka17, Barnacka18}. Galaxies may act as gravitational telescopes by means of their mass distributions. 
As a result of gravitational lensing, multiple  images of the same
background source (e.g., a quasar or a dual AGN system) may be observed, which can help uncover complex multi-wavelength structure at distances otherwise impossible to reach with current instruments \citep[e.g.][]{Congdon18}.

As discussed in \citet{Barnacka18}, high precision astrometric measurements have potential to reveal and elucidate diverse phenomena. Gravitationally lensed $\gamma-$ray flares were shown to occur 1.5 kpc from the nucleus in the blazar PKS~1830-211 at z=2.507 \citep{Barnacka15}.
For the lensed blazar B2~0218+35 at z=0.944, \citet{Barnacka16} showed that
the $\gamma$-ray flare was separated by at least 60 pc from the radio core. Direct measurement of optical to radio offsets at $z>1$ has been investigated by 
\citet{Spingola20}. The authors found an offset of 214$\pm$137 pc between the optical and radio emission of CLASS~B1608+656, and that optical and radio were co-spatial in CLASS~B0712+472 to within 17$\pm$42 pc. 
Also, the lensing magnification enabled to find that AGN jets and star forming regions are not spatially coincident in JVAS~B1938+666 at z=2.059 and MG~J0751+2716 at z=3.2 \citep{Spingola2020b}, and RX~1131-1231 at z=0.654 \citep{Paraficz2018}, possibly showing AGN feedback in action. 

Multi-wavelength studies are important for all these objectives; however, high energy wave bands are limited by the resolution
capability of current and even planned instrumentation.  In the X-ray band, the
state of the art angular resolution is provided by the \axaf\ observatory.  The \axaf\ precision of offset measurements is generally limited by the absolute celestial location aspect solution, which currently\footnote{https://cxc.harvard.edu/cal/ASPECT/celmon/} is 0\farcs8 to 90\% confidence for sources within 3\arcmin\ of the optical axis. At redshifts greater than a few tenths, 0\farcs8 corresponds to several kpc uncertainty. Gravitational lenses can be used to reduce this uncertainty by  up to two orders of magnitude, as we demonstrate herein for the dual AGN candidate MG~B2016+112. We thus move astrometric investigation of co-spatial emission to the high energy part of the electromagnetic spectrum. 

\vspace{0.5cm}
The paper is organized as follows. In Section~\ref{2016}, we describe properties of radio-loud gravitationally lensed quasar MG B2016+112. Section~\ref{Sec:xray_astrometry} describes our Bayesian method, while we report our results on the X-ray emission of MG B2016+112 in Section~\ref{Sec:results}. Finally, we discuss the results and present our conclusions in Section~\ref{Sec:discussion_conclusions}.

\section{MG B2016+112}\label{2016}
MG B2016+112 is a radio-loud source at redshift of 3.273 \citep{Lawrence1993} that is gravitationally lensed mainly by an early type galaxy and its faint satellite at redshift of 1.001 \citep{Schneider1986, Koopmans2002}. The lensing galaxies are part of a large cluster of galaxies \citep{Soucail2001}. At half-arcsec angular resolution, this system consists of three lensed images, separated by $\sim$3\farcs5.  Original X-ray detections by \citet{Hattori97} were interpreted as from the cluster, however \axaf\ observation by \citet{Chartas01} resolved three faint, point-like images consistent with the radio positions. Using the notation of \citet{Spingola19} we will call the X-ray images A1, B1 and C11. 

In the optical and radio bands image C is extended into a small arc \citep{Koopmans2002}, and  consists of the blending of the so-called ``merging images" (C11+C21) of a quadruply imaged system, but the two counter images were too faint to be detected in the radio with the VLBI observations of \citet{More2009} and \citet{Spingola19}. In fact, the lens mass model predicts these two images to be very close to images A1 and B1 and extremely faint (with flux density at 1.7 GHz of a few $\mu$Jy, \citealt{Spingola19}). Instead, images A and B are mirror images of the part of the source lying outside the caustic in the the doubly imaged region. At the radio wavelengths it is possible to spatially resolve on mas-scales all of the lensed images using VLBI observations. \citet{More2009} find that the lensed images are resolved into multiple compact and extended components with flat and steep radio spectra.

By comparing the position of these sub-components as measured with two VLBI observations separated by $\sim14$ years, \citet{Spingola19} found a significant positional change for four of them. 
The proper motions were of a few micro-arcseconds per year in the source plane for two of the radio components of the system MG B2016+112.  Such motion, once focused back in the source plane, suggested two possible scenarios for the nature of the source.
The first hypothesis consists of a single AGN source, seen at a small  viewing angle to the line of sight,  where the doubly-imaged jet is Doppler boosted (hence, the superluminal velocity) while the quadruply-imaged counter-jet is not boosted (hence, the subluminal velocity).
The other possibility is that of a dual AGN scenario, where there would be two jetted AGN (one doubly-imaged and one quadruply-imaged), seen under different viewing angles (one significantly Doppler boosted), separated by $\sim200$ pc, thus very close to be a physically bound binary AGN system. This scenario is supported by the detection of proper motion of the quadruply-imaged radio core, which is usually observed as a stationary component of the jet \citep[e.g.,][]{Boccardi17}, and the multi-wavelength photometric and spectroscopic  properties of the system (see Sec. 5.2 of \citealt{Spingola19} for a detailed discussion).  

To assess the nature of this system, it is, therefore, fundamental to investigate its X-ray properties. The presence of two X-ray sources, both with a flat photon index or intrinsic absorption, would be the smoking gun for confirming the dual AGN nature of MG B2016+112.

\begin{figure*}
\centering
\includegraphics[width=5.in]{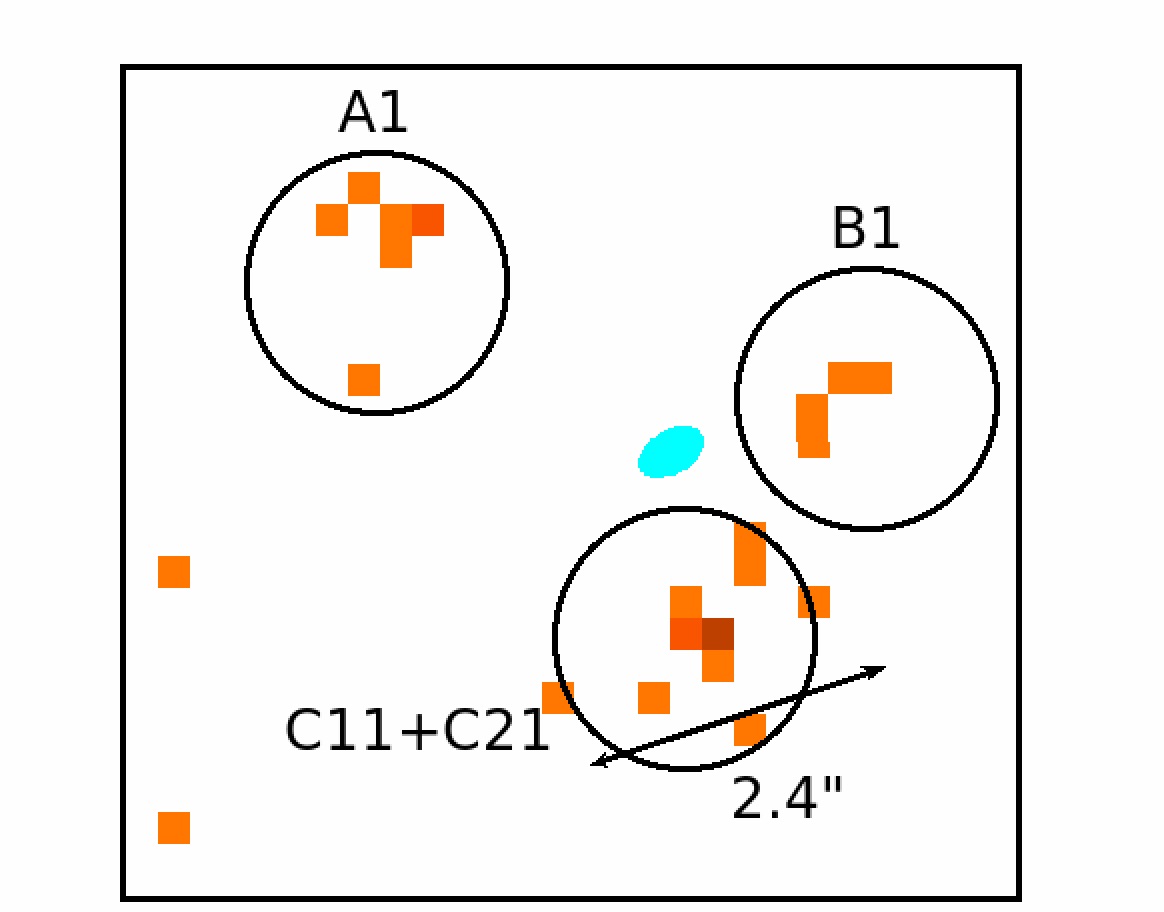}
\caption{MG B2016+112 X-rays from 0.5 to 7 keV,  binned in 0\farcs246 (1/2 ACIS pixel size) cells. Orange, red and brown are 1, 2, 3 X-rays per cell, respectively. Circles of 1\arcsec radius about A1, B1, and C11+C21 contain 7, 4, and 13 counts, respectively. The lensing galaxy D would be at the position of the cyan ellipse. The box covers 7\arcsec$\times$6\farcs5, the area containing the 28$\times$26  array of cells used to correlate to the data in Section~\ref{bayesanalysis} step 4. White areas have no measured counts.  \label{fig:xrayData}}
\end{figure*}

\section{X-ray Astrometry}\label{astrometry}
\label{Sec:xray_astrometry}
We present our X-ray analysis of the Chandra archival data in Section~\ref{analysis} and apply raytrace lensing analysis in Section~\ref{lensanalysis}. To further improve localization of the X-ray emission, in Section~\ref{bayesanalysis} we introduce and apply a method based on Bayesian prior constraints on the relative positions and intensities imposed by the model of the lens. 
\subsection{X-ray Analysis}\label{analysis}

We have re-analyzed the \axaf\ data of \citet[][ObsID 429]{Chartas01} using REPRO4, and the energy dependent subpixel event redistribution (EDSER) algorithm. Our 0.5 to 7 keV band image is shown in
Fig.~\ref{fig:xrayData}.  There are three distinct regions consistent with point sources in this very short 7.77~ks
observation. The regions are well separated in X-rays, and highly significant above the background of 0.01 count arcsec$^{-2}$.
The 7.77~ks observation yielded very few counts, 7, 4, and 13 for
images A, B, and C, 
respectively. With such poor statistics, independent estimates of the location of each image have $\sim\pm$0\farcs3 errors. 

\subsection{Raytrace Lensing Analysis}\label{lensanalysis}

We began by conducting inverse raytrace analysis following \cite{Spingola20}. 
Combining well reconstructed model of the lens and positions of the X-ray emission identified as the 3 images A1, B1, and C11
resulted in position errors of $\pm$80 to $\pm$100 mas for the location of the X-ray emission projected to the source plane. 
At the scale 7.7 pc per mas\footnote{We adopt ${\rm H}_{0}=67.8\rm\,km\,s^{-1}\,Mpc^{-1}$, $\Omega_{\rm
M}=0.308$ and $\Omega_{\rm \Lambda}=0.692$, \citep{Planck16}}, this is already
unparalled metric accuracy for an X-ray image at redshift
greater than 3. The two sources turn out to be 53 mas apart in the source plane, but with the large errors quoted.
\subsection{Bayesian Lensing Analysis}\label{bayesanalysis}

To further improve accuracy of localizing and resolving X-ray emission of MG B2016+112 we developed a new method based on the Bayesian approach. 
We use the Bayesian prior constraints on the relative position and intensities imposed by the model of the lens.
This approach improves the location of the X-ray sources relative to the radio sources by another order of magnitude. 
We use the maximum likelihood correlation technique as presented in \citet{Spingola21}, and summarized briefly here.

\begin{figure}
\centering
 \includegraphics[width=0.5\textwidth]{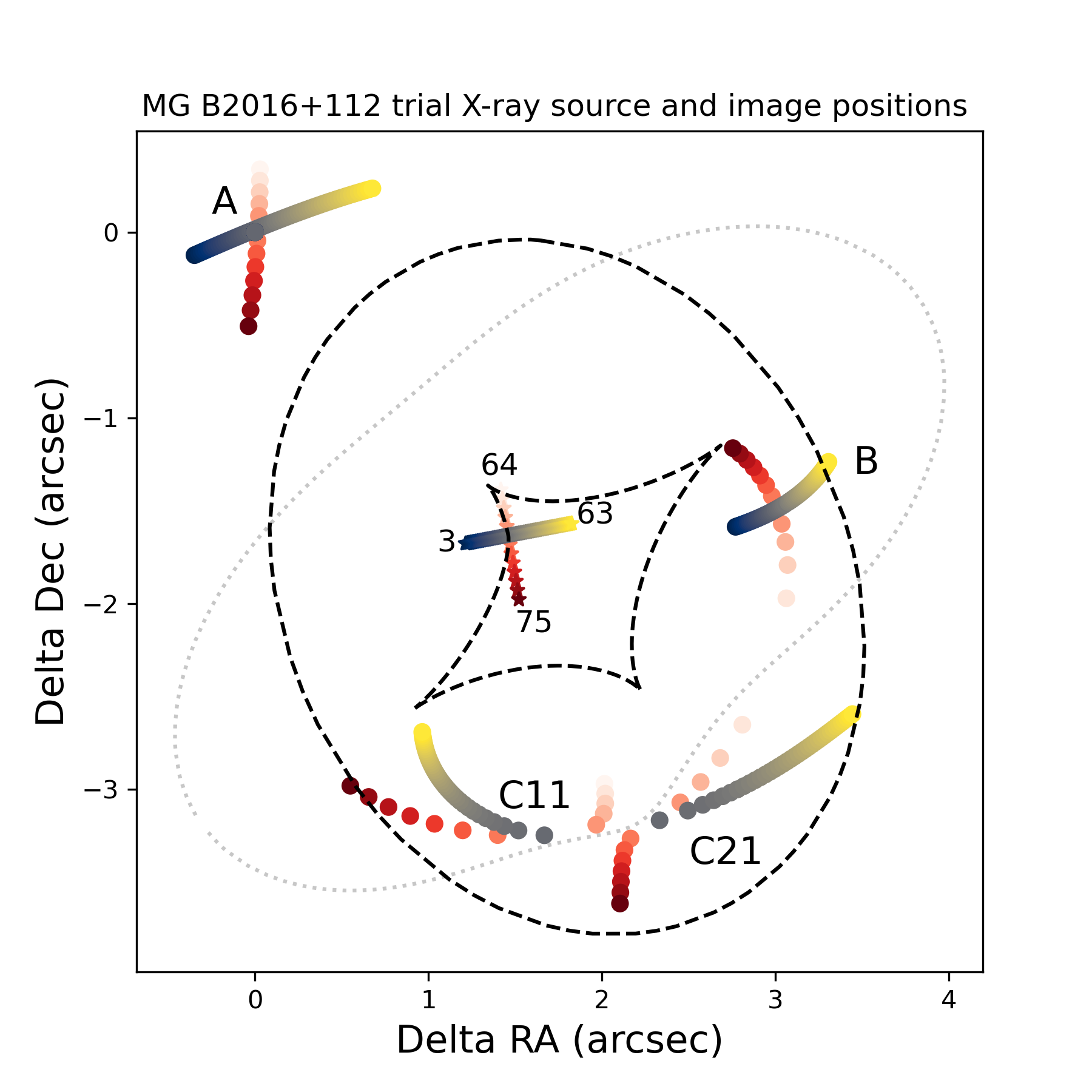}

\caption{Schematic of the trial source plane positions (stars) relative to the VLBI position of image A1, located at (0,0) arcsec. Source positions 3-63 are spaced by 10 mas perpendicular to the inner caustic (black dashed line), with positions 1 and 2 spaced 5 mas from position 26. Positions 64--75 are spaced 50 mas apart parallel to the caustic. The run of the resulting lensed images A1, B1, C11 and C21 (filled circles) are shown using the colour of the corresponding source. The dotted line represent the critical curve, while the black dashed line the caustic curves.\label{fig:trials}}
\end{figure}

{\bf Step 1:} For the \axaf\ response to a point source we run 1000 simulations with the actual aspect solution, dither, and total source flux from ObsID 429, and merge the results. Pileup is negligible for such a weak source.\\
{\bf Step 2:} At each of a grid of source plane positions we generate a high fidelity prediction for  the resulting lensed images designated A1, B1, C11, and C21 in \citet{Spingola19}. Trial source plane positions perpendicular to the caustic are spaced 10 mas apart. From SE to NW positions 1 -- 25 are outside the caustic (doubly imaged) and positions 26 -- 62 are inside the caustic and quadruply imaged (Fig.~\ref{fig:trials}). From NE to SW positions 63 -- 75 are inside and parallel to the caustic and spaced 50 mas apart. Since the amplification of the lens depends primarily on the distance from the caustic, our ability to distinguish positions parallel to the caustic is much coarser.
 For this grid of trial source positions we use the lensing mass model derived from the radio observations \citep{Spingola19} to predict the separations and magnifications of the four lensed images. \\
 {\bf Step 3:} We construct a model placing the simulated point source images from Step 1 at the predicted separations and with the predicted relative intensities. \\
{\bf Step 4:} We bin the observed X-ray data into a 28$\times$26 array of 0\farcs246 square pixels, with n$_i$ observed counts in each. \\
{\bf Step 5:} We raster our model of the four simulated lensed images plus background in two dimensions in steps of 24.6 mas and re-sort into the bins of the observed data array to predict the expected counts, $\lambda_i$, for each of the 728 data bins. \\
{\bf Step 6:} We compute the maximum likelihood for observing the counts in each bin based on Poisson statistics.  
\begin{equation}
C=-2P =-2 \sum_{i=1}^{728}\log(\lambda_i^{n_i} \exp(-\lambda_i)/n_i!)
\end{equation}
{\bf Step 7:} We linearly interpolate between raster coordinates to get the minimum value of -1/2 C (which we loosely refer to as the
``likelihood"), for each trial source position. 

We analyze two scenarios: 
{\bf Scenario 1:} One single source inside the caustic, or {\bf Scenario 2:}  two sources one inside and one outside caustic. For each scenario, we consider whether the relative fluxes of the lensed images are exactly as predicted from the lens model, or whether the amplitudes of each lensed image are free variables. In each case, we have only one interesting parameter per source: the number of the trial source position, or equivalently the position of the source perpendicular to or parallel to the local caustic in the source plane. We use Wilks theorem \citep{Wilks38, Cash79} that the differences of the likelihood function  
are distributed as $\chi^2$ with one degree of freedom  for each  interesting parameter. Numerically, we compute -1/2~C  and fit the 2-d raster position, the relative amplitudes of the lensed images (in cases where we allow them to vary from the lens model prediction), and the overall normalization of the simulated number of counts, all of which are  ``uninteresting parameters", to the observed total counts to give the minimum value for each trial source plane position.

\begin{figure}[t]
 \centering
 \includegraphics[width=3.in]{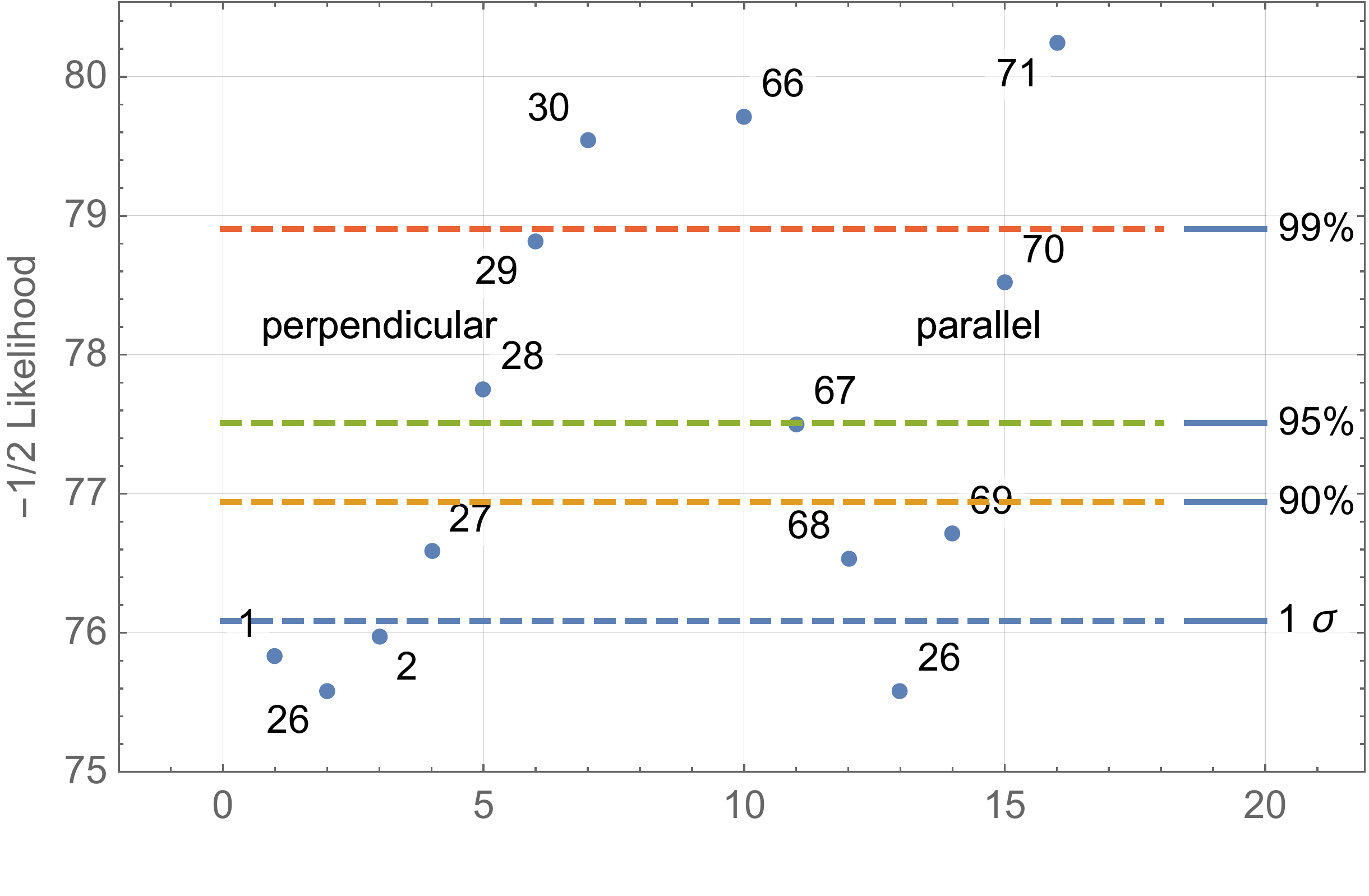}
\caption{Likelihood values for the case where one source produces all images, and the magnitudes of the lensed images are varied to give a minimum. Source positions 26 -- 30 are spaced by 10 mas perpendicular to the caustic, with positions 1 and 2 spaced 5 mas from position 26. Positions 66 -- 71 are spaced 50 mas parallel to the caustic. Dashed confidence levels are determined by $\chi^2$ for one degree of freedom relative to the minimum at position 26. \label{fig:onesource}}
\end{figure}


\section{Results}\label{Sec:results}

\subsection{Single source scenario\label{single}}
\begin{figure}
\includegraphics[width=3.in]{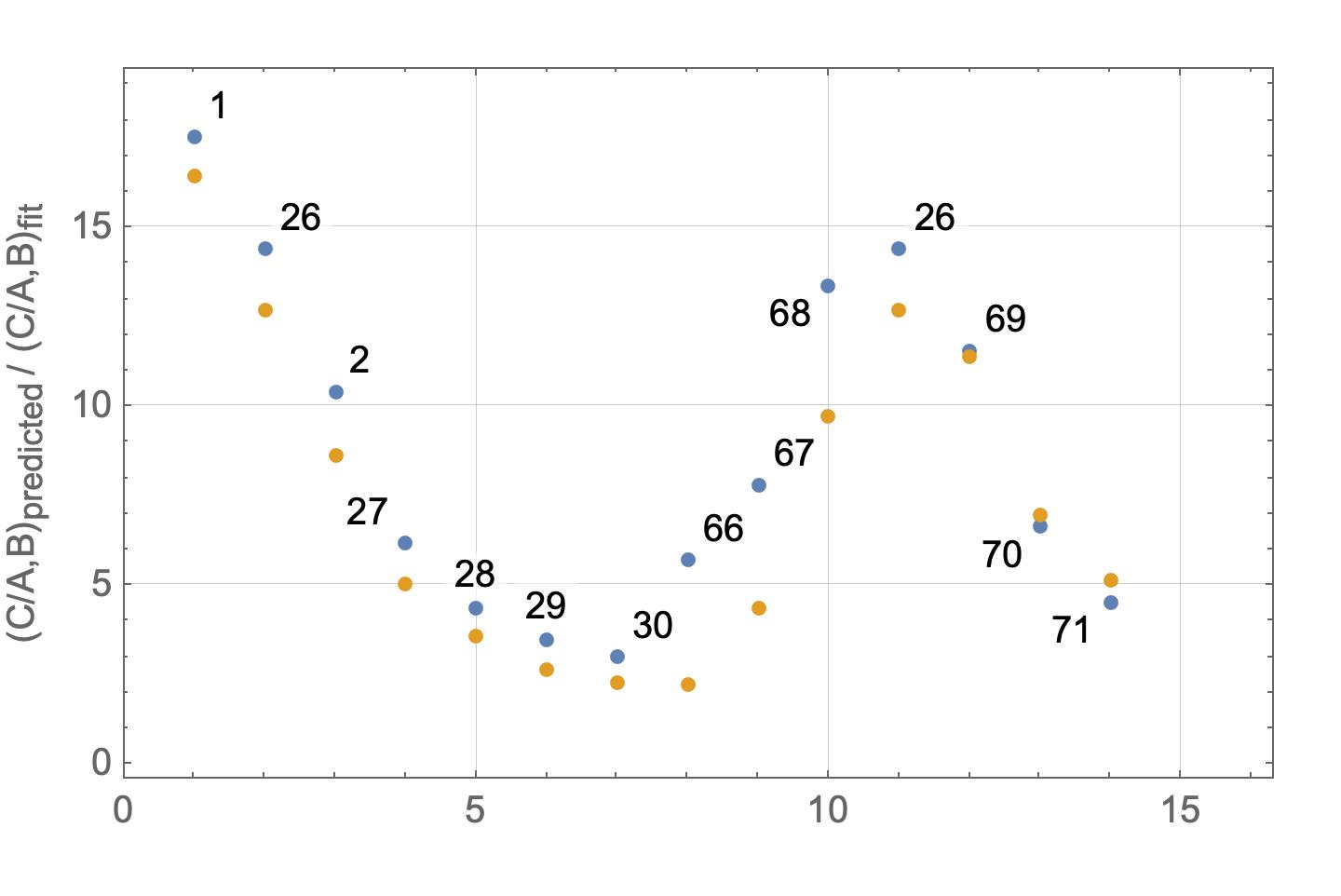}
\caption{For each position in Fig.~\ref{fig:onesource}, the expected ratio of the combined image C to image A (blue) or to image B (orange) as determined from the lens model, divided by the measured ratio of C to A (or to B) as determined by the maximum likelihood fit. Large values mean the C to A (B) observed ratio was much smaller than predicted, implying that A and B must be micro-lensed by a nearly equal, large factor.
\label{fig:ratios1}}
\end{figure}

Figure~\ref{fig:onesource} shows our results for the case where we assume there is only one X-ray source and allow the amplitudes of all lensed images to vary. Holding all amplitudes at the values of the predicted magnifications, we find a minimum likelihood value above 85, and thus can reject that hypothesis.   We plot -1/2 C for each of a series of trial source plane positions, as numbered. Position 26 is the inferred  position of source 4 epoch 1 from Table 4 of \citet{Spingola19}. To 90\% confidence one X-ray source is within 12 mas of the VLBI position perpendicular to the caustic, and within 70 mas parallel to the caustic. However there are some problems with this interpretation. Figure~\ref{fig:ratios1} shows the ratios of the expected magnifications of image C to image A (or to B) divided by the ratio of the numbers of counts which best fit image C to those best fitting image A (or B). The ratios near the maximum likelihood positions, \#26 and \#67-\#69 are much larger than 1. This is because the fitted counts in images A and B are much larger than expected, so the ratio C to A (or to B) is a smaller number in the denominator of the ratio that is plotted. Microlensing of any of the images is a reasonable explanation for what have long been observed as anomalous ratios among lensed images. In this case, however, both A and B are coincidentally both microlensed, and both by very nearly the same large factor, $\approx$ 14 to 18. Both the size and similarity of these factors are surprising and unlikely, noting that image A1 is much further from the lensing galaxy than image B1. Furthermore, from \citet{Koopmans02} Table 1, the summed flux ratio of images C to images A+B is 1.82$\pm$0.06, which we do not expect to be subject to extreme microlensing due to the size of the radio components. That ratio is consistent with the poorly determined ratio of 13/11 in the X-ray image.

The possibility that there is only one source, outside the caustic, can be rejected, since that would predict no X-ray flux at the position of C11+C21. The latter would then have to be an unrelated foreground or background source, for which the probability of being within 1\arcsec\ of the VLBI source is only of order 10$^{-5}$.

\subsection{Double source scenario\label{double}}
For the lens model to predict substantial counts as observed for images A and B, without extreme microlensing, we require a second X-ray source outside the caustic. This had been concluded from the VLBI observations \citep{More2009, Spingola19}, and we now establish that the X-rays also require a source inside and another outside the caustic.  We successively take 5 trial source positions spaced 50 mas apart perpendicular to and outside of the caustic. For each of those positions we compute the maximum likelihood as above (Sec. \ref{bayesanalysis}), pairing with all the positions inside the caustic. Since the amplification of the lens decreases rapidly outside the caustic to less than a factor of 2, we are much less sensitive to measuring the position of this outside source. 

\begin{figure*}
    \centering
   \includegraphics[width=4.in]{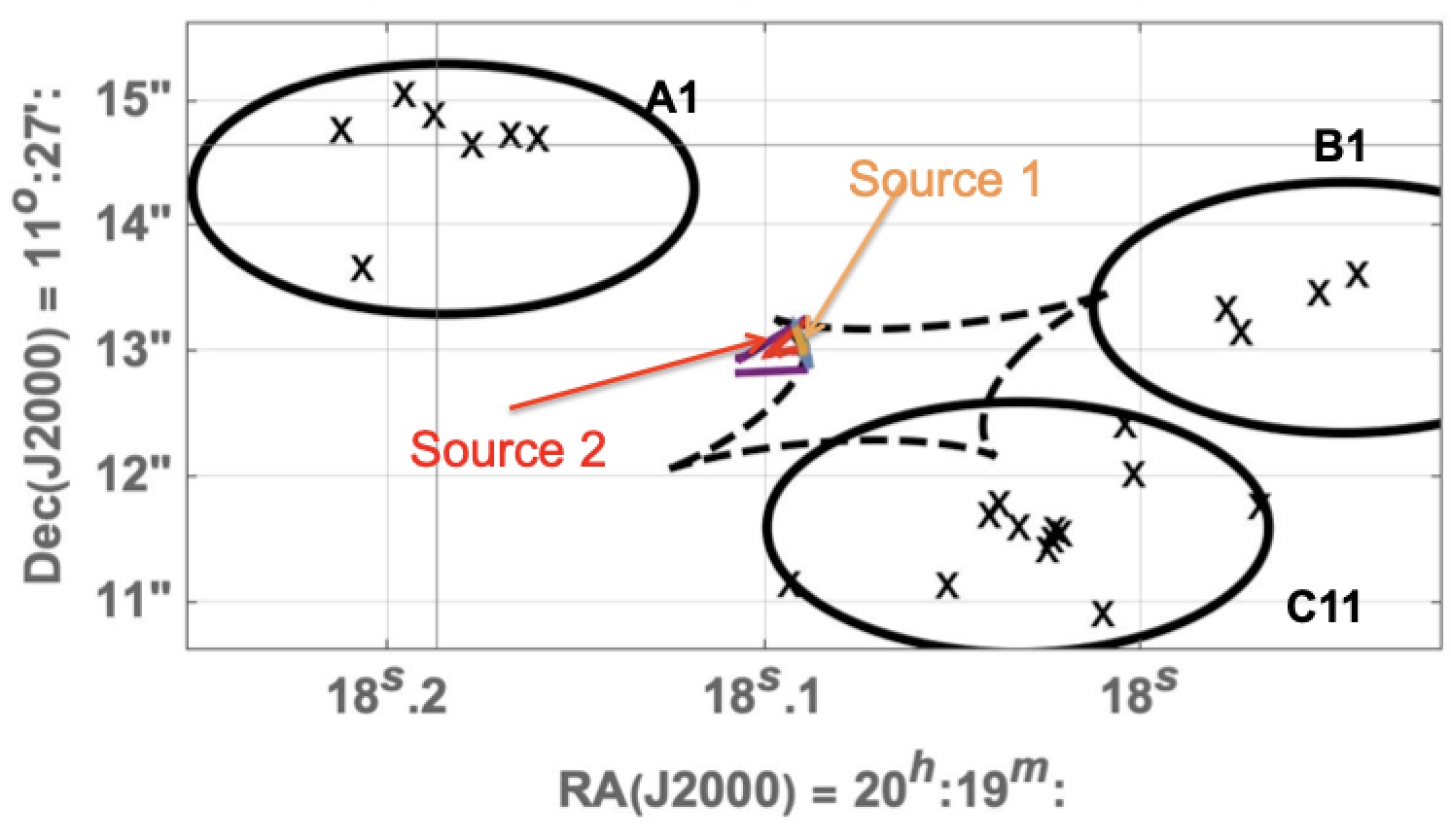}
\hspace{0.1in}
\includegraphics[width=2.in]{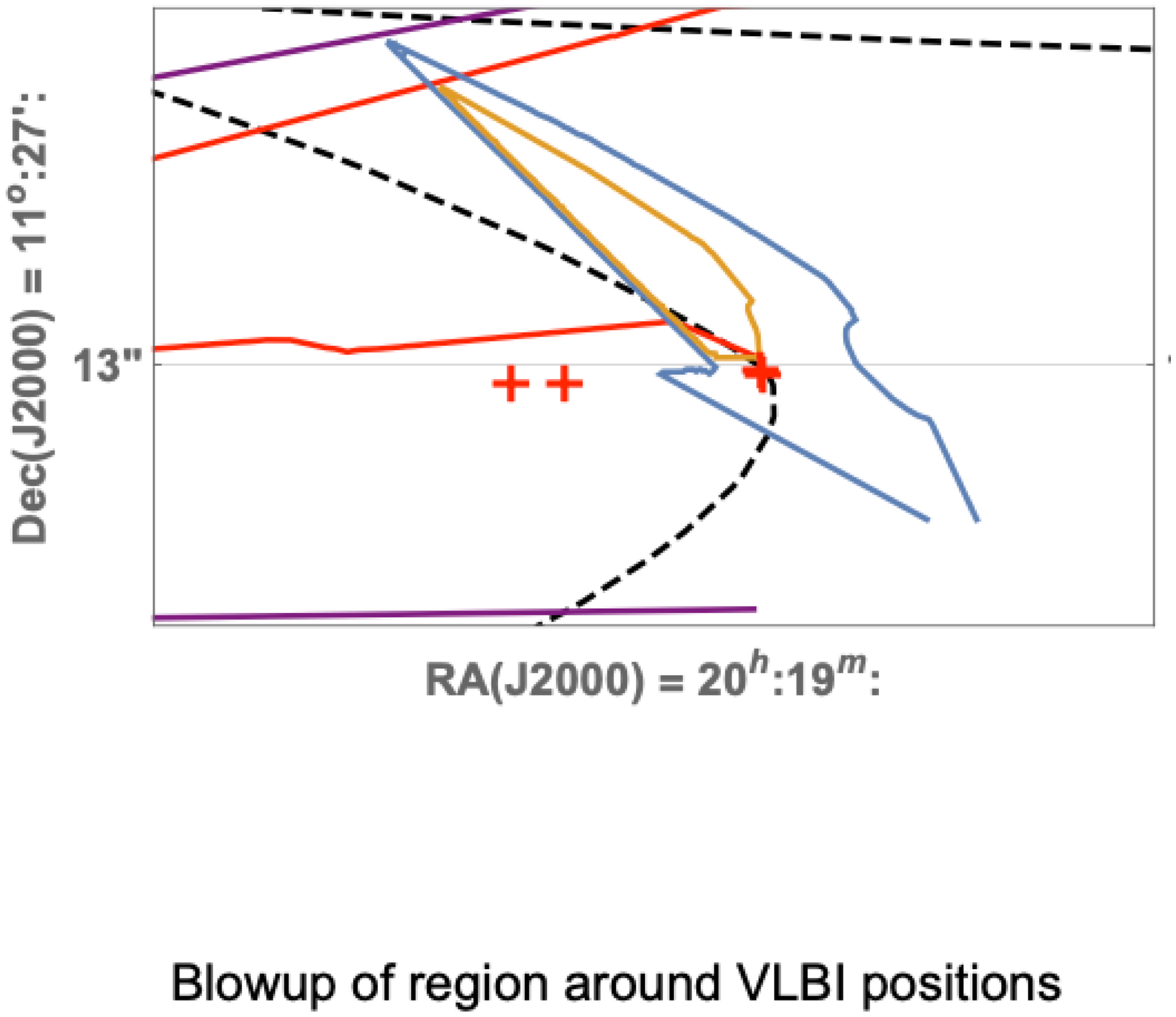}
   \caption{Left:  Celestial coordinates for the allowed locations of source 1 (inside the caustic, to lower RA) and source 2 (outside the caustic, to higher RA). Source 1 contributes predominantly to image C, and source 2  contributes only to images A and B. Contours enclose the two source positions to 68\%, and 90\% confidence, respectively,  (orange and blue for source  1, red and purple for source 2 in on-line version). The heavy dashed line shows the caustic. The "x's" are the locations measured for each of the 24 X-ray photons. They are associated with lensed images A1, B1, and C11 as shown by the heavy 1\arcsec\ circles (distorted in this projection). The heavier grid lines cross at the VLBI position of A1: 20$^h$19$^m$18.187$^s$ 11\degs27\arcmin14\farcs631. Right: Enlargement of the region around the VLBI positions, to better assess the constraints on the X-ray source locations. The red "+" signs are at the positions deduced for the four VLBI components in \citet{Spingola19}}
    \label{fig:twosources}
\end{figure*}

Allowing all the image amplitudes to vary results in flux ratios within a factor of 3 of the predicted magnifications, and gives the minimum values for our likelihood statistic. Freezing amplitudes at their predicted ratios gives just slightly higher values. These are statistically allowed and give smaller regions of allowed locations. We report results based on the variable image ratios as these are more conservative, and can be explained by reasonable microlensing and/or source variability.

Figure~\ref{fig:twosources} summarizes our position determinations for each of the two sources. We take a grid of the quantity -1/2 C, interpolated in RA and Dec coordinates relative to the position deduced for source 4, epoch 1, in Table 4 of \citet{Spingola19}. For source 1 (source 2) we marginalize over the likelihood of source 2 (source 1) to construct confidence contours considering two interesting degrees of freedom. To within a full range of $\pm 1\sigma$ (alternately, of $\pm$90\%) confidence we locate source 1 within 10  (20) mas perpendicular to the caustic, and 170 (315) mas parallel to the caustic. For source 2  our analysis is much less restrictive  because the lower amplification changes the image positions by smaller angular distances than can be distinguished for such few X-ray counts. With 1$\sigma$ confidence, source 2 is restricted to be  within 160 mas perpendicular to the caustic, and within 200 mas parallel to the caustic. To 90\% confidence we limit source 2 within 500 mas parallel to the caustic, but do not restrain its perpendicular extent outside the caustic.  
Those are the full range of uncertainty, conventionally those uncertainties would be reported as $\pm$ 1/2 the numbers quoted above. 

We have measured the location of source 1 within a region 77$\times$1300 pc, to 68\% confidence. This is unprecedented metric accuracy in X-ray astronomy for an object at such a large redshift. 

 \section{Discussion and conclusions} \label{Sec:discussion_conclusions}

\subsection{The Nature of the Two X-ray Sources\label{nature}}

We fit the 24 photons from all the lensed images to a power law plus galactic absorption of 0.15$\times$10$^{22}$ cm$^{-2}$. The fit gives a power law photon index $\Gamma=0.82\pm0.39$ where dn/dE=E$^{-\Gamma}$. 
This is an extremely flat value for an AGN. A better fit is obtained by allowing intrinsic absorption at the source redshift. We find (17$\pm 14)\times$10$^{22}$ cm$^{-2}$, and a  photon power law index of  1.52$\pm$0.46 that would be reasonable for a quasar. The total energy flux measured from 0.5 to 7 keV would be 3.1$\times$10$^{-14}$ erg cm$^{-2}$ s$^{-1}$, or 1.4 and 1.7  $\times$10$^{-14}$ erg cm$^{-2}$ s$^{-1}$ for source 2 (sum of images A and B) and source 1 (image C), respectively. The 11 photons in images A and B are distributed between 818 and 5753 eV, while the 13 photons in image C are distributed from 1294 to 4313 eV. This hints that source 1 has a steeper spectrum and more intrinsic absorption, but the sparse statistics allow all the spectra to be identical. 
If the two X-ray sources are actually at the positions of the VLBI components given by \citet{Spingola19}, their magnifications would be 30 for source 1 and 1.8 for source 2. This would make their intrinsic 0.5 -- 7 keV  luminosities 2.6$\times$10$^{43}$ and 4.2$\times$10$^{44}$ erg s$^{-1}$ in the source frame, respectively. We caution that the magnification gradient is very steep through the allowed locations of source 1, which therefore could have an intrinsic luminosity several times more or even 10 times less. 

The X-ray results are completely consistent with observing two gravitationally lensed AGN. The usual interpretation of radio quiet AGN is that the X-ray emission is from a hot corona, very near to the supermassive black hole. The enhanced X-ray to optical ratios for radio loud AGN have been suggested to be due to beamed X-ray emission \citep[e.g.,][]{Zamorani1981,Worrall1987}.
A dual AGN interpretation is consistent with the narrow emission line spectra observations of \citet{Yamada2001} that showed different ionization properties for A+B vs. for C. Both AGN would be relatively low luminosity type II quasars, consistent with the large intrinsic X-ray absorption. 

While the VLBI observations also favor a dual AGN \citep{Spingola19}, we can not rule out interpretation as a single X-ray AGN, where we separately image the core and its extended X-ray jet. The intrinsic ratio of 20 for source 2 to source 1 would be consistent with observation of kpc X-ray jets, which found a median of about 50 for a set of low redshift quasars \citep{Marshall2018}. However, such a small contribution from a jet would not explain the enhanced ratio of X-ray to optical emission from radio loud quasars. Observations of the X-ray spectra of the three images will be crucial to establish a dual AGN. 

\subsection{The future of X-ray astrometry using gravitational lensing }
MGB 2016+112 was only the fourth gravitational lens system to be discovered and only the second to be a radio source \citep{Lawrence1984}. It may prove to be an important ``Rosetta Stone" for high energy astrophysics. Most exciting is the prospect that we are imaging an X-ray jet on the scale of pc from a supermassive black hole at large redshift. Alternately, establishing a dual or possibly binary supermassive black hole at z$>$3 will have implications for the frequency of such systems in the early Universe.

\citet{Spingola21} and the present study are the first to exploit the power of gravitational lenses as telescopes for the measurement of X-ray source positions. With a high fidelity lens mass model we used the predicted constraints on the lensed images to locate by direct observation that an X-ray source was within a few 10's of pc of VLBI emission in MG B2016, a quasar at redshift 3.273. Many X-ray lens systems have far greater photon statistics, and will be a crucial resource for astrometric study of X-ray structure at large redshift.

Future surveys with the next generation of radio and optical telescopes, such as the Square Kilometre Array (SKA), \textsl{Euclid} and the ``Vera C. Rubin" Observatory, will allow the discovery of $\sim10^5$
strong lensing systems \citep{Oguri2010, Collett2015, McKean2015}. Among them, there will be rare background sources, such as dual and binary AGN candidates. If quadruply imaged, like MG B2016+112, these systems will provide an unprecedented opportunity to investigate the final SMBH merging stages at high spatial resolution and at cosmological distances. Nevertheless, it will still be necessary to confirm the dual/binary AGN nature of these sources. The X-ray properties are crucial for this aim, e.g. by comparing the sources photon index and intrinsic absorption. Our novel method can be applied to X-ray follow-up observations of the most promising dual and binary AGN candidates that will be discovered in the future lensing suveys.






\end{document}